\RequirePackage{lineno}
\documentclass[aps,twocolumn,showpacs,byrevtex,prl,reprint,nofootinbib]{revtex4-1}
\usepackage{xspace}

\usepackage{graphicx}
\usepackage{dcolumn}
\usepackage{bm}
\usepackage{rotating}
\usepackage{color}
\usepackage{verbatim} 
\usepackage{multirow}
\usepackage{amsmath}
\usepackage{relsize}
\usepackage{ulem}
\usepackage{mathrsfs}
\begin{document}

\title{\boldmath Measurement of the absolute branching fraction of the inclusive decay $\Lambda_c^+ \to K_S^0X$ }
\author{
\small
M.~Ablikim$^{1}$, M.~N.~Achasov$^{10,d}$, P.~Adlarson$^{64}$, S. ~Ahmed$^{15}$, M.~Albrecht$^{4}$, A.~Amoroso$^{63A,63C}$, Q.~An$^{60,48}$, ~Anita$^{21}$, Y.~Bai$^{47}$, O.~Bakina$^{29}$
% ...
\\
\vspace{0.2cm}
(BESIII Collaboration)\\
\vspace{0.2cm} {\it
$^{1}$ Institute of High Energy Physics, Beijing 100049, People's Republic of China\\
% ...
$^{4}$ Bochum Ruhr-University, D-44780 Bochum, Germany\\
% ...
$^{10}$ G.I. Budker Institute of Nuclear Physics SB RAS (BINP), Novosibirsk 630090, Russia\\
% ...
$^{15}$ Helmholtz Institute Mainz, Johann-Joachim-Becher-Weg 45, D-55099 Mainz, Germany\\
% ...
$^{21}$ Indian Institute of Technology Madras, Chennai 600036, India\\
% ...
$^{29}$ Joint Institute for Nuclear Research, 141980 Dubna, Moscow region, Russia\\
% ...
$^{47}$ Southeast University, Nanjing 211100, People's Republic of China\\
$^{48}$ State Key Laboratory of Particle Detection and Electronics, Beijing 100049, Hefei 230026, People's Republic of China\\
% ...
$^{60}$ University of Science and Technology of China, Hefei 230026, People's Republic of China\\
% ...
$^{63}$ (A)University of Turin, I-10125, Turin, Italy; (B)University of Eastern Piedmont, I-15121, Alessandria, Italy; (C)INFN, I-10125, Turin, Italy\\
$^{64}$ Uppsala University, Box 516, SE-75120 Uppsala, Sweden\\
% ...
$^{d}$ Also at the Novosibirsk State University, Novosibirsk, 630090, Russia\\
% ...
}
}

\date{\today}

\begin{abstract}

We report the first measurement of the absolute branching fraction of the inclusive decay $\Lambda_c^+ \to K_S^0X$.
The analysis is performed using an $e^+e^-$ collision data sample corresponding to an integrated luminosity of 567~pb$^{-1}$
taken at $\sqrt{s}$ = 4.6~GeV with the BESIII detector.
Using eleven Cabibbo-favored $\bar{\Lambda}_c^-$ decay modes and the double-tag technique,
this absolute branching fraction is measured to be $\mathcal{B}(\Lambda_c^+ \to K_S^0X)=(9.9\pm0.6\pm0.4)\%$,
where the first uncertainty is statistical and the second systematic.
The relative deviation between the branching fractions for the inclusive decay and the observed exclusive decays is $(18.7\pm8.3)\%$,
which indicates that there may be some unobserved decay modes with a
neutron or excited baryons in the final state.

\end{abstract}


\maketitle

The lightest charmed baryon $\Lambda_c^+$ was first observed in $e^+e^-$ annihilation at the Mark~II experiment~\cite{PRL44-10}.
Hadronic $\Lambda_c^+$ decays offer an ideal platform to understand both strong and weak interactions.
Most branching fractions (BFs) of $\Lambda_c^+$ decays were previously measured relative to the BF of $\Lambda_c^+\to p K^-\pi^+$~\cite{pdg}. 
In recent years, the BESIII experiment reported a series of absolute measurements of exclusive decays of the $\Lambda_c^+$
baryon~\cite{PRL115-221805, PRL116-052001, PRL117-232002, PLB767-42, PLB783-200, PRD99-032010, PRL118-112001, PLB772-388}.
The precision of BFs for the known decay modes was significantly improved and some new decay modes were observed.
Using the statistical isospin model~\cite{PRD97-116015}, it is estimated that 
about 90\% of the $\Lambda_c^+$ decay modes are now known.  
Measurements of the BFs for inclusive decays of the $\Lambda_c^+$ baryon are important to understand its decay mechanisms
and indicate the size and type of unmeasured decays by comparing with the BFs for the corresponding exclusive decays.

The Cabibbo-favored (CF) decays of charmed mesons have been well studied~\cite{pdg}.
However, the information of the CF decays of charmed baryons is relatively limited.
The $\Lambda_c^+$ CF decays are dominantly modes involving $\Lambda$, $\Sigma$ and $\bar{K}$ in the final state.
According to the statistical isospin model,
the total BF of the observed and extrapolated CF decays of $\Lambda_c^+$ baryon is $(83.2\pm4.9)\%$~\cite{PRD97-116015}.
Measurements of the BF of the inclusive decays will help to characterize $\Lambda_c^+$ CF decays.
Recently, BESIII measured the absolute inclusive BF ${\mathcal B}(\Lambda_c^+ \to \Lambda X) = (38.2^{+2.8}_{-2.2}\pm0.9$)\%~\cite{PRL121-062003},
which appears to be larger than the total observed and extrapolated BFs for exclusive $\Lambda$ decays ($31.7\pm1.4$)\%~\cite{PRD97-116015}.
The total BF of exclusive $\bar{K}^0/K^0$ decays of $\Lambda_c^+$ is estimated to be $(22.4\pm0.9)\%$ by the statistical isospin model~\cite{PRD97-116015},
as listed in Table~\ref{tab_exp}, while the total observed BF for decays to $\bar{K}^0/K^0$ only sum to ($16.1\pm0.8$)\%.
Determining the absolute BF of inclusive $\Lambda_c^+$ decays to $\bar{K}^0/K^0$
will help to quantify the missing decay modes and test the predicted BFs of decay modes extrapolated by the statistical isospin model.

In this paper, we measure the absolute BF of the inclusive decay of the $\Lambda_c^+$ to $K_S^0$ ($\Lambda_c^+ \to K_S^0X$) for the first time,
where $X$ indicates all possible particle combinations.
This analysis uses 567~pb$^{-1}$ of data~\cite{CPC39-093001} collected at 
the center-of-mass energy $\sqrt{s} = 4.6$~GeV with the BESIII detector.
The measurement is performed using the double-tag (DT) technique~\cite{dtag},
since there is no additional hadrons accompanying $\Lambda_c^+\bar{\Lambda}_c^-$ pair produced at this energy.
First, the $\bar{\Lambda}_c^-$ baryons are reconstructed with exclusive hadronic decay modes which are called the single-tag (ST) modes.
Then the $\Lambda_c^+ \to K_S^0X$ mode is reconstructed in the $\bar{\Lambda}_c^-$ recoiling side, called the signal mode or the DT mode.
The ST $\bar{\Lambda}_c^-$ baryons are reconstructed including the following eleven hadronic decay modes:
$\bar{p}K_S^0$, $\bar{p}K^+\pi^-$, $\bar{p}K_S^0\pi^0$, $\bar{p}K_S^0\pi^+\pi^-$, $\bar{p}K^+\pi^-\pi^0$, $\bar{\Lambda}\pi^-$, $\bar{\Lambda}\pi^-\pi^0$,
$\bar{\Lambda}\pi^-\pi^+\pi^-$, $\bar{\Sigma}^0\pi^-$, $\bar{\Sigma}^-\pi^0$, and $\bar{\Sigma}^-\pi^+\pi^-$, with a total BF of $(35.0\pm0.7)\%$.
Throughout this paper, charge-conjugate modes are implicitly assumed unless explicitly stated.

\begin{table}[htbp]
\begin{center}
\caption{Observed and extrapolated BFs for exclusive $\bar{K}^0/K^0$ decays of $\Lambda_c^+$ CF decays~\cite{pdg, PRD97-116015}.
Here, observed BFs are referred from Particle Data Group (PDG)~\cite{pdg} and extrapolated BFs are referred from Ref.~\cite{PRD97-116015}.
BFs of the $\bar{K}^0/K^0$ decay modes are obtained by doubling those quoted for $K_S^0$ decay modes.
The total uncertainty is obtained as the sum in quadrature.}
\label{tab_exp}
\begin{tabular*}{0.48\textwidth}{@{\extracolsep{\fill}}lclc}
\hline
\hline
Mode            & Value (\%)  &  Mode  &  Value (\%) \\
\hline
\multicolumn{2}{c}{Observed BF} & \multicolumn{2}{c}{Extrapolated BF} \\
\hline
$p\bar{K}^0$              & 3.18$\pm$0.16   & $n\bar{K}^0\pi^+\pi^0$         & 3.07$\pm$0.16 \\
$p\bar{K}^0\pi^0$         & 3.94$\pm$0.26   & $p\bar{K}^0\pi^0\pi^0$         & 1.36$\pm$0.07 \\
$p\bar{K}^0\pi^+\pi^-$    & 3.20$\pm$0.24   & $n\bar{K}^0\pi^+\pi^+\pi^-$    & 0.14$\pm$0.09 \\
$n\bar{K}^0\pi^+$         & 3.64$\pm$0.50   & $p\bar{K}^0\pi^+\pi^-\pi^0$    & 0.22$\pm$0.14 \\
$p\bar{K}^0\eta$          & 1.60$\pm$0.40   & $n\bar{K}^0\pi^+\pi^0\pi^0$    & 0.10$\pm$0.06 \\
$\Lambda K^+\bar{K}^0$    & 0.57$\pm$0.11   & $p\bar{K}^0\pi^0\pi^0\pi^0$    & 0.03$\pm$0.02 \\
                          &                 & $(\Sigma K)^+\bar{K}^0$        & 0.68$\pm$0.34 \\
                          &                 & $\Xi^0K^0\pi^+$                & 0.62$\pm$0.06 \\
\hline
Total & $16.1\pm0.8$ & Total & $6.3\pm0.4$ \\
\hline
Total & \multicolumn{3}{c}{22.4$\pm$0.9} \\
\hline
\hline
\end{tabular*}
\end{center}
\end{table}

The BESIII detector is described in detail in Ref.~\cite{bibbes3}.
It has an effective geometrical acceptance of 93\% of 4$\pi$.
The cylindrical core of the BESIII detector consists of a small-cell, helium-based (40\% He, 60\% C$_3$H$_8$) multi-layer drift chamber (MDC),
a plastic scintillator time-of-flight system (TOF), a CsI(Tl) electromagnetic calorimeter (EMC),
and a muon system containing resistive plate chambers in the iron return yoke of a 1~T superconducting solenoid.
The momentum resolution for charged tracks is 0.5\% at a momentum of 1~GeV/$c$.
Charged particle identification (PID) is accomplished by combining the energy loss ($dE/dx$) measurements in the MDC and flight times in the TOF.
The photon energy resolution at 1 GeV is 2.5\% in the barrel and 5\% in the end caps.

A Monte Carlo (MC) simulation based on {\footnotesize GEANT}4~\cite{bibgeant4} includes
the geometric description of the BESIII detector and its response.
We generate high-statistics MC samples to study the background and estimate the detection efficiencies; 
initial-state radiation (ISR)~\cite{EAVS} and final-state radiation~\cite{FSR} are also included in the MC simulation.
$\Lambda_c^+\bar{\Lambda}_c^-$ pairs, $D_{(s)}^{(*)}\bar{D}_{(s)}^{(*)}X$ production, ISR production of $\psi$ states, and continuum $q\bar{q}$ processes
are simulated with generic MC samples generated using the KKMC generator~\cite{kkmc}. 
The known decay modes are simulated with {\footnotesize EVTGEN}~\cite{evtgen} using BFs taken from PDG~\cite{pdg},
and the remaining unknown decays are simulated with the {\footnotesize LUNDCHARM} model~\cite{PRD62-034003}.

Charged tracks are detected in MDC.
For prompt tracks, the polar angle ($\theta$) is required to satisfy $|\cos\theta|<0.93$,
and the point of closest approach to the interaction point (IP) is required to be less than 10~cm in the beam direction and less than 1~cm in the transverse plane.
Secondary tracks used to reconstruct $K_S^0$ or $\bar{\Lambda}$ candidates 
are subject to different IP requirements as detailed below.  
Particle identification (PID) for charged tracks combining
the measurements of the energy loss $dE/dx$ in the MDC and the flight time information
is employed to calculate a likelihood $\mathcal{L}(h)$ for each hadron ($h=p, K$, or $\pi$) hypothesis.
Protons, kaons and pions are identified by requiring that the likelihood for the given hypothesis is larger than for both of the other two hypotheses.

Photon candidates are reconstructed by clustering electromagnetic calorimeter (EMC) crystal energies.
The deposited energy is required to be greater than 25 MeV in the EMC barrel region ($|\cos\theta|<0.80$) 
and 50 MeV in the EMC end cap region ($0.86<|\cos\theta|<0.92$).
To eliminate showers from charged particles, the angle between the photon and the nearest charged track is required to be greater than $20^\circ$.
Timing requirements are used to suppress electronic noise and energy deposits in the EMC unrelated to the event.
$\pi^0$ candidates are reconstructed from photon pairs with an invariant mass in the range $0.115<M_{\gamma\gamma}<0.150$~GeV/$c^2$.
A mass-constrained fit to the $\pi^0$ nominal mass~\cite{pdg} is performed to improve the momentum resolution.

$K_S^0$ and $\bar{\Lambda}$ candidates are reconstructed by combining pairs of oppositely charged tracks ($\pi^+\pi^-$ for $K_S^0$ and $\bar{p}\pi^+$  for $\bar{\Lambda}$)
satisfying $|\cos\theta|<0.93$ for the polar angle. The distance to the IP in the beam direction is required to be within 20~cm.
No distance constraints in the transverse plane are required.
Charged pions from these decays are not subjected to the PID requirement, while proton PID is applied in order to improve signal significance. 
The two charged tracks are constrained to originate from a common decay vertex by requiring the $\chi^2$ of the vertex fit to be less than 100.
Furthermore, the decay vertex is required to be separated from the IP by a distance of at least twice the uncertainty of the vertex fit.
To select $K_S^0$, $\bar{\Lambda}$, $\bar{\Sigma}^0$, and $\bar{\Sigma}^-$,
the invariant mass of $\pi^+\pi^-$, $\bar{p}\pi^+$, $\bar{p}\pi^+\gamma$ and $\bar{p}\pi^0$ are 
required to be within (0.487, 0.511)~GeV/$c^2$, (1.111, 1.121)~GeV/$c^2$, (1.179, 1.203)~GeV/$c^2$ and (1.176, 1.200)~GeV/$c^2$, respectively.

For the ST modes $\bar{p}K_S^0\pi^0$, $\bar{p}K_S^0\pi^+\pi^-$ and $\bar{\Sigma}^-\pi^+\pi^-$,
background events containing a $\bar{\Lambda}$ are rejected by vetoing candidate events with $M(\bar{p}\pi^+)$ in the interval (1.110, 1.120)~GeV/$c^2$.
$K_S^0$ backgrounds for the ST modes $\bar{\Lambda}\pi^-\pi^+\pi^-$, $\bar{\Sigma}^-\pi^0$ and $\bar{\Sigma}^-\pi^+\pi^-$
are suppressed by requiring $M(\pi^+\pi^-)$ or $M(\pi^0\pi^0)$ to be outside of (0.480, 0.520)~GeV/$c^2$.
To remove $\bar{\Sigma}^-$ background in the ST mode $\bar{p}K_S^0\pi^0$,
candidates within the range $1.170<M(\bar{p}\pi^0)<1.200$~GeV/$c^2$ are excluded.

\begin{figure}[hbtp]
\centering
\includegraphics[width=0.48\textwidth]{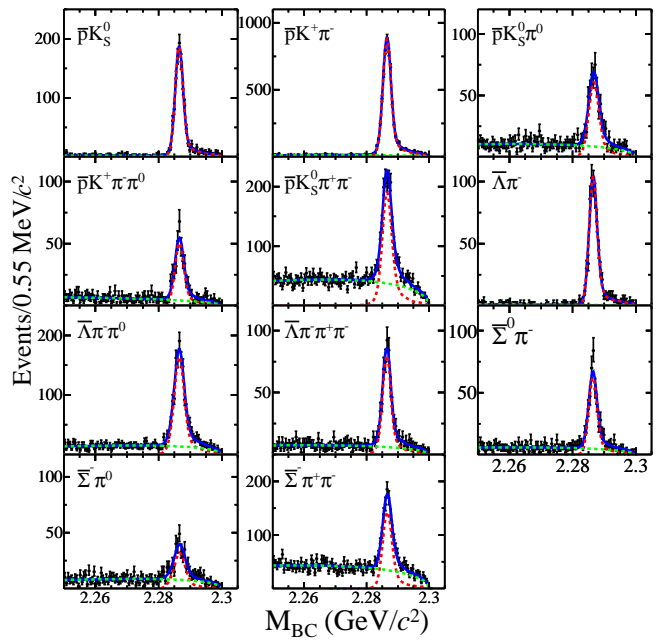}
\caption{Fits to the distributions of $M_{\rm BC}$ in data sample for different ST $\bar{\Lambda}_c^-$ modes,
where the black dots with error bars are data, the blue lines are the fit results, 
the dashed red lines are signal shapes, and the dashed green lines are background shapes.
}
\label{fig00}
\end{figure}

The quantities $M_{\rm BC}=\sqrt{E^2_{\rm beam}-|\vec{p}_{\bar{\Lambda}_c^-}|^2}$ and $\Delta E=E_{\bar{\Lambda}_c^-}-E_{\rm beam}$
are used to identify ST $\bar{\Lambda}_c^-$ candidates,
where $E_{\rm beam}$ is the beam energy and $E_{\bar{\Lambda}_c^-}$ and $\vec{p}_{\bar{\Lambda}_c^-}$ are energy and momentum of the $\bar{\Lambda}_c^-$ candidate.
To improve the signal purity, $|\Delta E|$ requirements corresponding to about three times the resolutions are imposed on $\bar{\Lambda}_c^-$ candidates; 
details are given in Table~\ref{tab00}.  
If there is more than one candidate per ST mode, the one with minimum $|\Delta E|$ is chosen.
The $\bar{\Lambda}_c^-$ signals are clearly visible in the $M_{\rm BC}$ distributions of the eleven tag modes, as shown in Fig.~\ref{fig00}. 
Peaking backgrounds are negligible according to MC studies~\cite{topo}.
Unbinned maximum likelihood fits to $M_{\rm BC}$ distributions are used to determine the ST yields for each tag mode, 
where the signal shape is described by the MC-simulated shape convolved with a Gaussian function to better match the resolution found in data,
and the background shape is described by an ARGUS function~\cite{argus}.
The resultant ST yields in the signal region $2.282 < M_{\rm BC}< 2.300$~GeV/$c^2$ and the corresponding detection efficiencies are listed in Table~\ref{tab00}.

\begin{table}[htbp]
\begin{center}
\caption{Requirements on $\Delta E$, ST yields in data ($N_i^{\rm tag}$ ), ST ($\epsilon_i^{\rm tag}$) and DT ($\epsilon_i^{\rm tag,sig}$) efficiencies for the tag mode $i$.
Uncertainties on $N$ are statistical only, while uncertainties on efficiencies are due to the MC statistics.
The quoted efficiencies do not include any BFs of subsequent decays.}
\begin{tabular*}{0.48\textwidth}{@{\extracolsep{\fill}}lcccc}
\hline
\hline
Mode                           & $\Delta E$ (MeV)  & $N_i^{\rm tag}$ &  $\epsilon_i^{\rm tag}$ (\%) & $\epsilon_i^{\rm tag,sig}$ (\%) \\
\hline
$\bar{p}K_S^0$                 & ($-20, 19$)  &  1222$\pm$37 & 55.3$\pm$0.2  &   26.6$\pm$0.4   \\
$\bar{p}K^+\pi^-$              & ($-20, 15$)  &  6024$\pm$85 & 49.2$\pm$0.1  &   24.9$\pm$0.2   \\
$\bar{p}K_S^0\pi^0$            & ($-30, 20$)  &   498$\pm$29 & 18.9$\pm$0.1  &   8.7$\pm$0.2    \\
$\bar{p}K_S^0\pi^+\pi^-$       & ($-20, 20$)  &   376$\pm$24 & 15.5$\pm$0.1  &   7.1$\pm$0.2    \\
$\bar{p}K^+\pi^-\pi^0$         & ($-30, 20$)  &  1544$\pm$57 & 16.1$\pm$0.1  &   7.8$\pm$0.1    \\
$\bar{\Lambda}\pi^-$           & ($-20, 20$)  &   693$\pm$30 & 42.1$\pm$0.2  &  22.4$\pm$0.4    \\
$\bar{\Lambda}\pi^-\pi^0$      & ($-30, 20$)  &  1362$\pm$47 & 14.1$\pm$0.1  &   6.9$\pm$0.1    \\
$\bar{\Lambda}\pi^-\pi^+\pi^-$ & ($-20, 20$)  &  569$\pm$30  & 11.5$\pm$0.1  &   5.4$\pm$0.1    \\
$\bar{\Sigma}^0\pi^-$          & ($-20, 20$)  &  438$\pm$26  & 25.2$\pm$0.1  &  12.0$\pm$0.4    \\
$\bar{\Sigma}^-\pi^0$          & ($-50, 30$)  &  291$\pm$32  & 23.0$\pm$0.2  &  12.1$\pm$0.4    \\
$\bar{\Sigma}^-\pi^+\pi^-$     & ($-30, 20$)  &  1111$\pm$50 & 23.7$\pm$0.1  &  11.9$\pm$0.2    \\
\hline
\hline
\end{tabular*}
\label{tab00}
\end{center}
\end{table}

\begin{figure}[hbtp]
\centering
\includegraphics[width=0.24\textwidth]{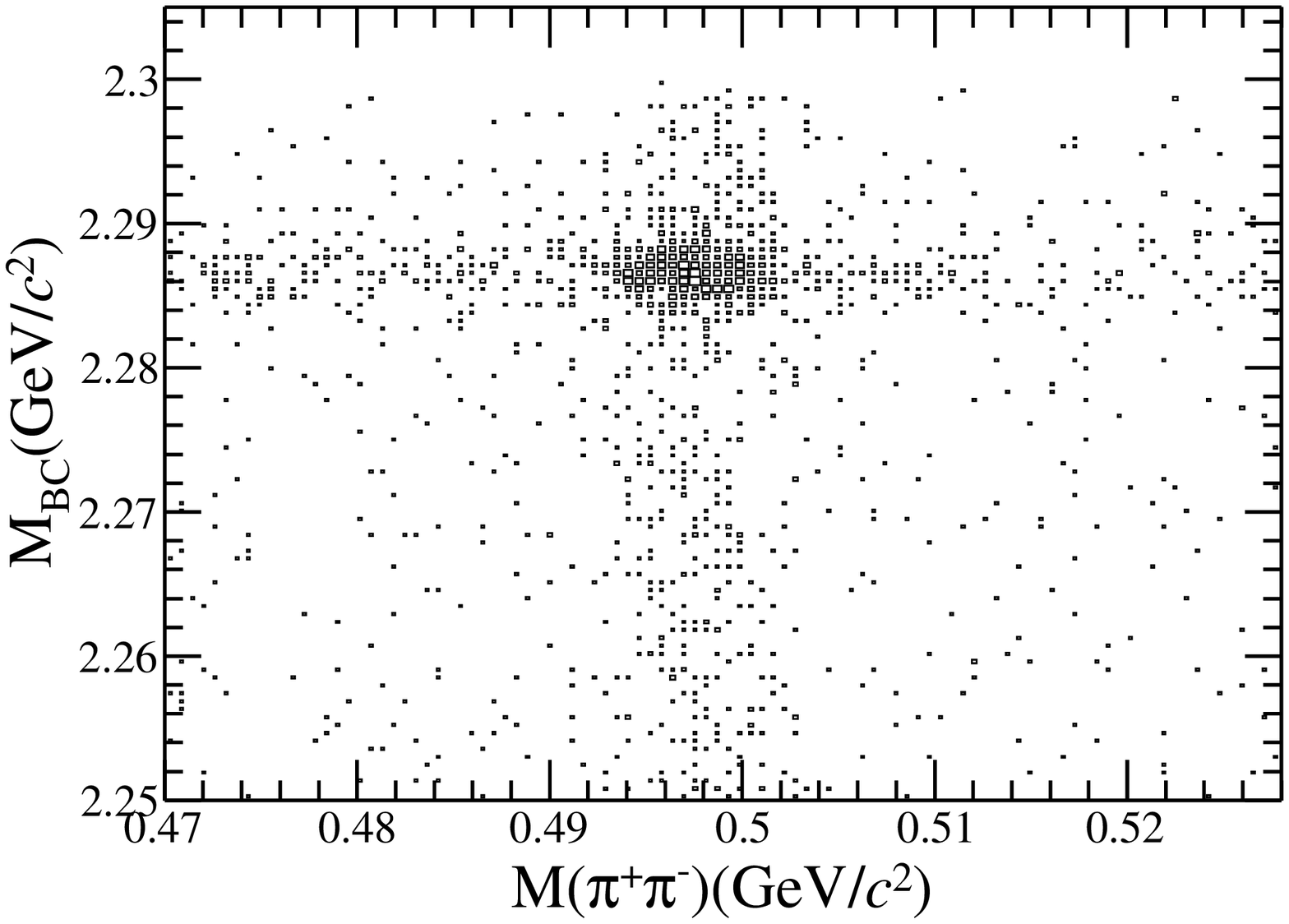}
\put(-100,70){\bf\boldmath (a)} 
\includegraphics[width=0.24\textwidth]{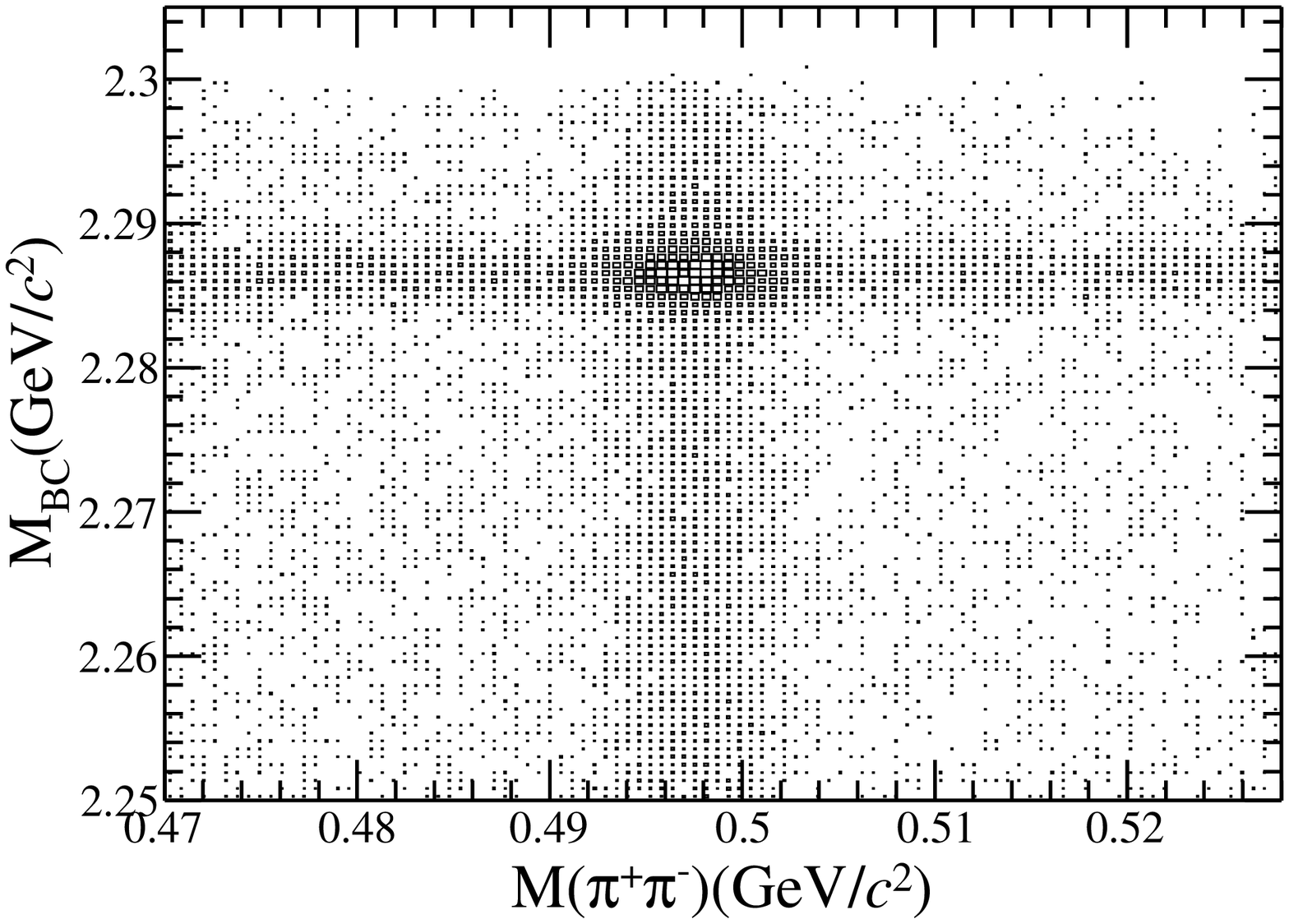}
\put(-100,70){\bf\boldmath (b)} \\
\includegraphics[width=0.24\textwidth]{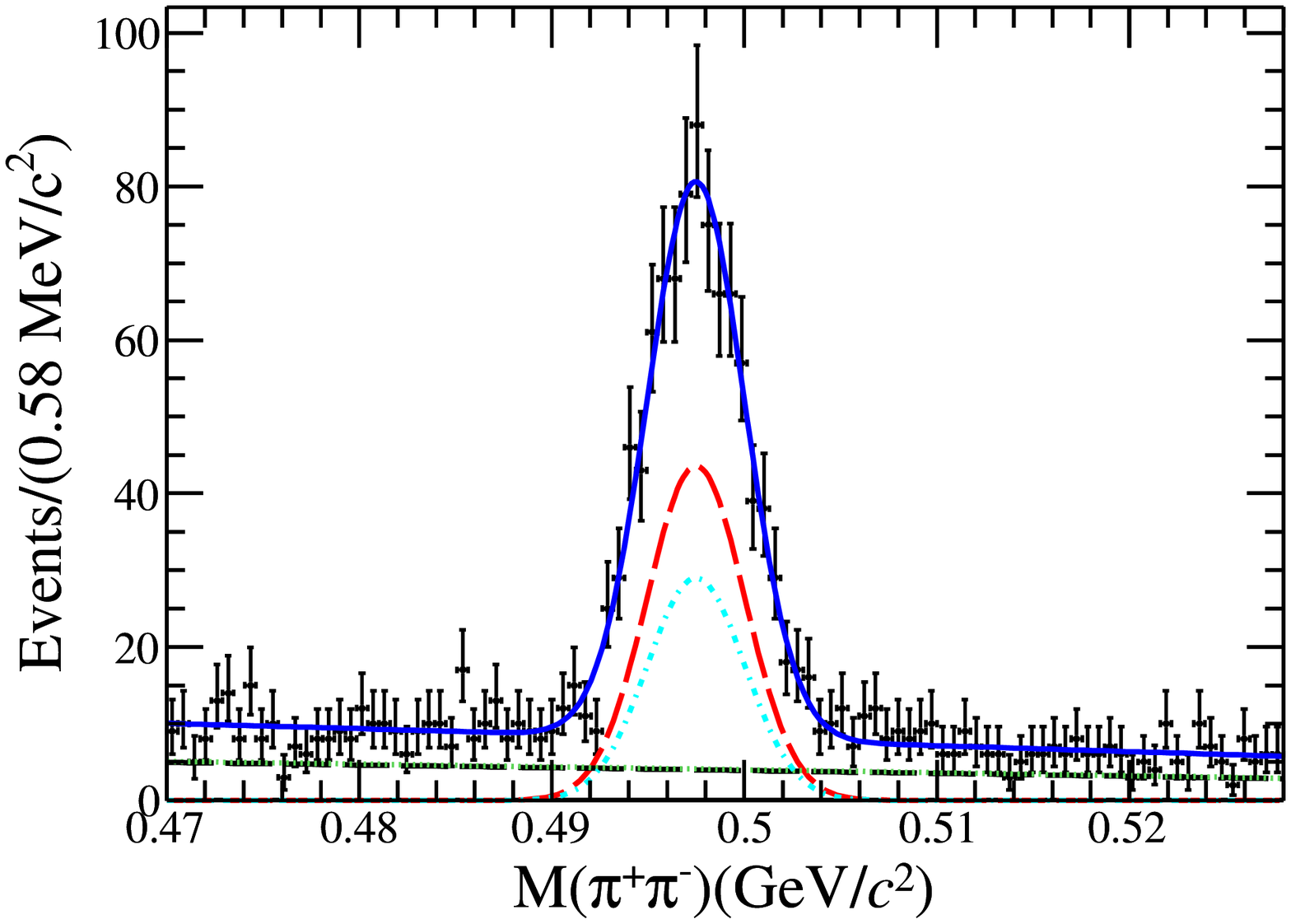} 
\put(-100,70){\bf\boldmath (c)}
\includegraphics[width=0.24\textwidth]{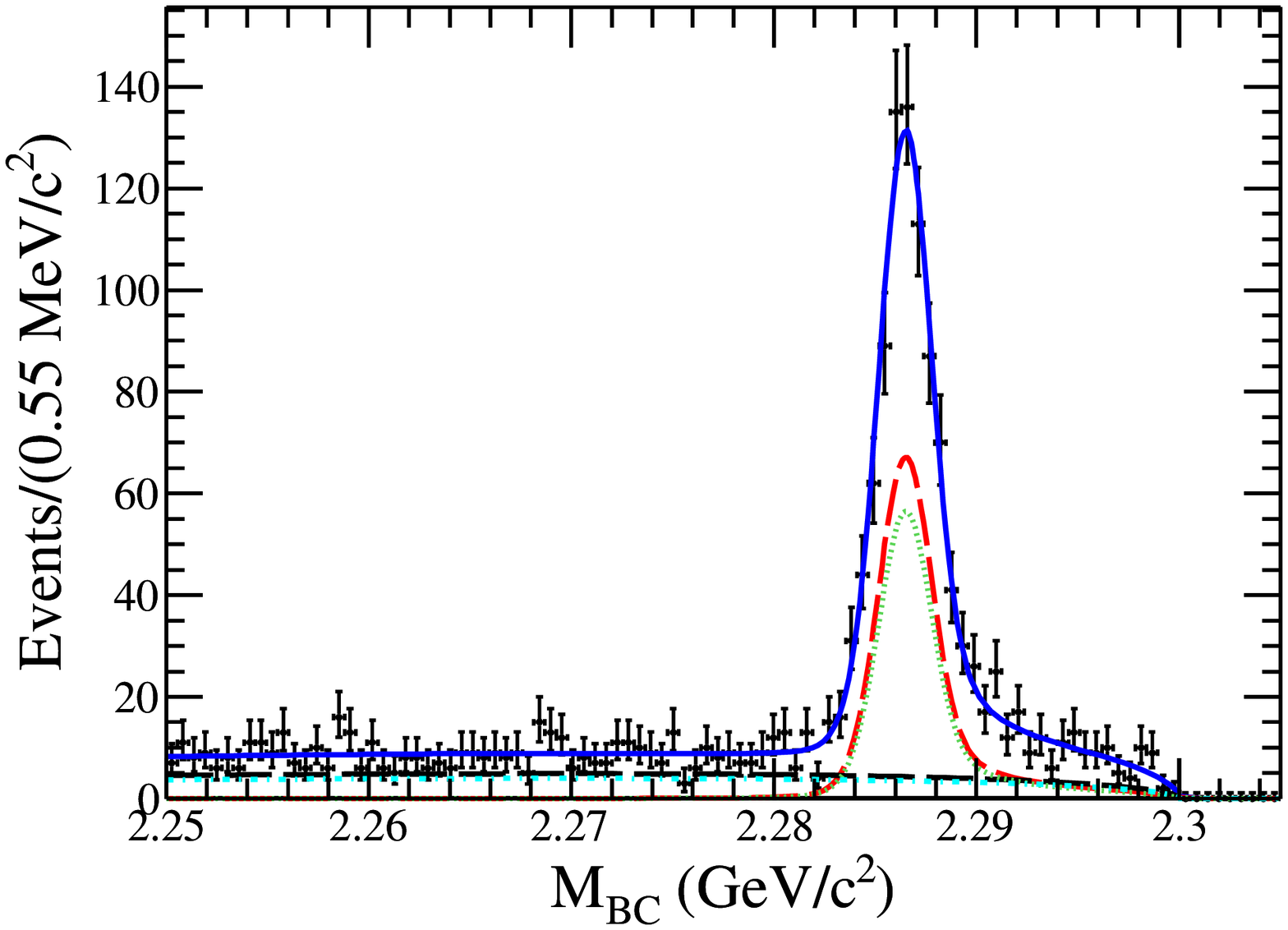} 
\put(-100,70){\bf\boldmath (d)} 
\caption{ [(a) and (b)] Distributions and [(c) and (d)] projections of $M_{\rm BC}$ versus $M(\pi^+\pi^-)$ of the DT candidate in (a) data and (b) the 2D fit result,
where the black dots with error bars are data, the blue solid curves are the fit results,
the red-dashed lines are signal function, the black-dashed lines are background neither peaking in the $M(\pi^+\pi^-)$ distribution nor the $M_{\rm BC}$ distribution,
the green-dotted lines are background peaking around the $\bar{\Lambda}_c^-$ mass in the $M_{\rm BC}$ distribution,
and the cyan-dash-dotted lines are background peaking around the $K_S^0$ mass in the $M(\pi^+\pi^-)$ distribution. 
}
\label{fig01}
\end{figure}

We select $K_S^0$ candidates among the remaining tracks on the recoiling side of the tagged $\bar{\Lambda}_c^-$.
The selection criteria of $K_S^0$ are the same as those used in the ST $\bar{\Lambda}_c^-$ selection.
If there is more than one $K_S^0$ candidate, the one with the minimum vertex fit $\chi^2$ is selected for further analysis.
Figure~\ref{fig01}~(a) shows the distribution of $M_{\rm BC}$ versus the invariant mass of $\pi^+\pi^-$ pairs, $M(\pi^+\pi^-)$, of the accepted candidates for all eleven tag modes.
There is a clear $\Lambda_c^+ \to K_S^0X$ signal in the intersection of the $K_S^0$ and the ST $\bar{\Lambda}_c^-$ signal bands.
A two-dimension (2D) fit to the distribution of $M_{\rm BC}$ versus $M(\pi^+\pi^-)$ is performed to determine the signal yield, as shown in Fig.~\ref{fig01}.
The signal function is the product of the $\bar{\Lambda}_c^-$ signal function and $K_S^0$ signal function.
There are three kinds of background:
the background peaking neither in the $M_{\rm BC}$ distribution nor in the $M(\pi^+\pi^-)$ distribution is described by 
the product of $\bar{\Lambda}_c^-$ background function and $K_S^0$ background function;
the background peaking around the $\bar{\Lambda}_c^-$ mass in the $M_{\rm BC}$ distribution is described by
the product of $\bar{\Lambda}_c^-$ signal function and $K_S^0$ background function;
the background peaking around the $K_S^0$ mass in the $M(\pi^+\pi^-)$ distribution is described by
the product of $\bar{\Lambda}_c^-$ background function and $K_S^0$ signal function.
The $\bar{\Lambda}_c^-$ signal is described by the MC-simulated shape convolved with a Gaussian function, while background is ARGUS function.
The $K_S^0$ signal and background functions are described by a Gaussian function and a first-order polynomial, respectively.
The signal yield is fitted to be $478\pm27$, where the uncertainty is statistical.

The absolute BF $\mathcal{B}^{\rm sig} = {\mathcal B}(\Lambda_c^+ \to K_S^0X)$  is determined by 
\begin{equation}
\label{eq3}
\mathcal{B}^{\rm sig}=\frac{N^{\rm sig}}{\mathcal{B}(K_S^0\to \pi^+\pi^-)\cdot \sum_{i}N_i^{\rm tag}\cdot \epsilon_i^{\rm tag,sig}/\epsilon_i^{\rm tag}},
\end{equation}
where $\epsilon_i^{\rm tag,sig}$ is the DT efficiency for the tag mode $i$, as listed in Table~\ref{tab00}.
The absolute BF of $\Lambda_c^+ \to K_S^0X$ is calculated to be $\mathcal{B}(\Lambda_c^+ \to K_S^0X) = (9.9\pm0.6)\%$,
the uncertainty is statistical only.
The reliability of the analysis method used in this work has been validated by analyzing the generic MC sample.

Systematic uncertainties from the ST side mostly cancel in the BF measurement with the DT method.
Other systematic uncertainties for measuring $\mathcal{B}(\Lambda_c^+ \to K_S^0X)$
are described below and summarized in Table~\ref{tab01}. 

We refer to the systematic uncertainty for $\sum N_i^{\rm tag}\cdot \epsilon_i^{\rm tag,sig}/\epsilon_i^{\rm tag}$ as ST-related systematic uncertainty.
The systematic uncertainty of the ST yields ($N_i^{\rm tag}$) is studied by altering the signal shape, fitting range,
and end point of the ARGUS function. The uncertainty due to limited MC statistics is taken as the uncertainty of the ST and DT efficiencies
($\epsilon_i^{\rm tag}$ and $\epsilon_i^{\rm tag,sig}$). The total relative ST-related systematic uncertainty is calculated to be 1.2\%.
The systematic uncertainty of the $K^0_S$ reconstruction is determined to be 1.5\% by
studying control samples of $J/\psi\to K^{*\mp}K^{\pm}$ and $J/\psi\to\phi K_S^0K^{\pm}\pi^{\mp}$ and weighting over the momentum of the $K^0_S$~\cite{PRD92-112008}.
The systematic uncertainty for $\mathcal{B}(K_S^0\to \pi^+\pi^-)$ is 0.1\% from PDG~\cite{pdg}.
The systematic uncertainty of the signal yield is estimated by altering the $K_S^0$ signal function, background function and the 2D fit range,
The relative changes (3.4\%) in the BF are taken as systematic uncertainties.
Assuming no correlations between sources, the total systematic uncertainty is obtained as the sum in quadrature.

\begin{table}[htbp]
\begin{center}
\caption{Systematic uncertainties in the measurement of the BF of $\Lambda_c^+ \to K_S^0X$.}
\begin{tabular*}{0.48\textwidth}{@{\extracolsep{\fill}}lccc}
\hline
\hline
Source                           & Uncertainty (\%) \\
\hline
  ST related                  & 1.2 \\
  $K_S^0$ reconstruction      & 1.5 \\
  $\mathcal{B}(K_S^0\to \pi^+\pi^-)$ & 0.1 \\
  Signal yield & 3.4 \\
\hline
Total &  3.9 \\
\hline
\hline
\end{tabular*}
\label{tab01}
\end{center}
\end{table}

In summary, the absolute BF of the inclusive decay $\Lambda_c^+ \to K_S^0X$ is measured 
for the first time by using an $e^+e^-$ data sample of 567~pb$^{-1}$
taken at $\sqrt{s} = 4.6$~GeV with the BESIII detector.
The result is $\mathcal{B}(\Lambda_c^+ \to K_S^0X) = (9.9\pm0.6\pm0.4)\%$,
where the first uncertainty is statistical and the second systematic.
The BF of the inclusive decay $\Lambda_c^+ \to \bar{K}^0/K^0X$ is $(19.8\pm1.2\pm0.8\pm1.0)$\%
where the third uncertainty of $\pm$5\% is included to account for possible differences between
$\mathcal{B}(\Lambda_c^+ \to  K_S^0X)$ and $\mathcal{B}(\Lambda_c^+ \to  K_L^0X)$~\cite{PLB349-363},
which is consistent with calculations with the statistical isospin model within 1.3$\sigma$.
The relative BF deviation of $(18.7\pm8.3)\%$ between the inclusive $\bar{K}^0/K^0$ decay and the observed exclusive decays of $\Lambda_c^+$,
can be addressed by the extrapolated exclusive decays of $\Lambda_c^+$ listed in Table~\ref{tab_exp}.
Experimentally, only one decay mode involving a neutron in the final state was observed at BESIII~\cite{PRL118-112001}.
More decay modes involving neutrons or hyperons in the final states can be 
experimentally pursued,
especially decays with a large BF, e.g. $\Lambda_c^+\to n\bar K^0\pi^+\pi^0$ whose BF is calculated to be $(3.07\pm0.16)$\% by the statistical isospin model.
Recently, the BF of $\Lambda_c^+\to \Xi^0K^0\pi^+$ was calculated to be $(8.70\pm1.70$)\% by the SU(3) flavor symmetry model~\cite{PRD99-073003},
while it is only $(0.62\pm0.06)$\% in the statistical isospin model. Measuring the BF of $\Lambda_c^+\to \Xi^0K^0\pi^+$ will test these two models.

\par

The BESIII collaboration thanks the staff of BEPCII and the IHEP computing center for their strong support.
This work is supported in part by National Key Basic Research Program of China under Contract No. 2015CB856700; 
Chinese Academy of Science Focused Science Grant;
National 1000 Talents Program of China;
National Natural Science Foundation of China (NSFC) under Contracts Nos.
11905225, 11775230, 11935018, 11625523, 11635010, 11735014, 11822506, 11835012, 11935015, 11935016, 11521505, 11425524, 11605042, 11961141012;
the Chinese Academy of Sciences (CAS) Large-Scale Scientific Facility Program;
Joint Large-Scale Scientific Facility Funds of the NSFC and CAS under Contracts Nos. U1732263, U1832207; 
CAS Key Research Program of Frontier Sciences under Contracts Nos. QYZDJ-SSW-SLH003, QYZDJ-SSW-SLH040;
100 Talents Program of CAS;
INPAC and Shanghai Key Laboratory for Particle Physics and Cosmology;
ERC under Contract No. 758462;
German Research Foundation DFG under Contracts Nos. Collaborative Research Center CRC 1044, FOR 2359;
Istituto Nazionale di Fisica Nucleare, Italy;
Ministry of Development of Turkey under Contract No. DPT2006K-120470;
National Science and Technology fund;
STFC (United Kingdom);
The Knut and Alice Wallenberg Foundation (Sweden) under Contract No. 2016.0157;
The Royal Society, UK under Contracts Nos. DH140054, DH160214; 
The Swedish Research Council; 
U. S. Department of Energy under Contracts Nos. DE-FG02-05ER41374, DE-SC-0010118, DE-SC-0012069.

\end{document}